**Title:** A Systematic Review of Mutations Associated with Isoniazid Resistance Points to Lower Diagnostic Sensitivity for Common Mutations and Increased Incidence of Uncommon Mutations in Clinical Strains of *Mycobacterium tuberculosis*

**Running Title:** Systematic Review of mutations Associated with Isoniazid Resistance

**Authors:** Siavash J. Valafar [1*]

**Author Affiliation:** [1] Department of Biology, University of California in Irvine, Irvine, CA

* Corresponding Author name and contact: Siavash J. Valafar, Department of Biology, University of California in Irvine, Irvine, California 92617. Email: svalafar@uci.edu.

# ABSTRACT

Molecular testing is rapidly becoming integral to the global tuberculosis (TB) control effort. Uncommon mechanisms of resistance can escape detection by these platforms and lead to the development of Multi-Drug Resistant (MDR) strains. This article is a systematic review of published articles that reported isoniazid (INH) resistance-conferring mutations between September-2013 and December-2019. The aims were to catalogue mutations associated with INH resistance, estimate their global prevalence and co-occurrence, and their utility in molecular diagnostics. The genes commonly associated with INH resistance, *katG*, *inhA*, *fabG1*, and the intergenic region *oxyR'-ahpC* were considered in this review. In total, 52 articles were included describing 5,632 INH$^R$ clinical isolates from 31 countries. The three most frequently mutated loci continue to be *katG*315 (4,100), *inhA*-15 (786), and *inhA*-8 (105). However, the diagnostic value of *inhA*-8 is far lower than previously thought, only appearing in 25 (0.4%) INH$^R$ isolates that lacked a mutation at the first two loci. Importantly, of the four *katG* loci recommended by the previous systematic review for diagnostics, only *katG*315 was observed in our INH$^R$ isolates. This indicates continued evolution and regional differences in INH resistance. We have identified 58 loci (common to both systematic reviews) in three genomic regions as a reliable basis for molecular diagnostics. We also catalogue mutations at 49 new loci associated with INH resistance. Including all observed mutations provides a cumulative sensitivity of 85.1%. The most disconcerting is the remaining 14.9% of isolates that harbor an unknown mechanism of resistance, will escape molecular detection, and likely convert to MDR-TB, further complicating treatment. Integrating the information cataloged in this and other similar studies into current diagnostic tools is essential for combating the emergence of MDR-TB. Exclusion of this information will lead to an "unnatural" selection which will result in eradication of the common but propagation of the uncommon mechanisms of resistance, leading to ineffective global treatment policy and a need for region-specific regiments. Finally, the observance of many low-frequency resistance-conferring mutations point to an advantage of platforms that consider regions rather than specific loci for detection of resistance.

**INTRODUCTION**

Tuberculosis (TB) is the most prevalent infectious diseases to date, with an estimated ten million new cases in 2018. It is also the infectious disease with highest mortality, recently surpassing HIV/AIDS. One of the challenges in global TB control is the emergence of drug resistance. Globally, the World Health Organization (WHO) estimates a total of 186,772 Rifampicin (RIF) and Multidrug Resistant (MDR) (resistant to RIF and isoniazid [INH]) TB cases in 2018, a number that has been on the rise in spite of declining total global TB cases. Treatment success rate for MDR-TB is low (56%, globally), making its accurate diagnosis and prevention, when possible, a critical piece of global TB control. Traditionally culture-based methods are used to determine resistance. Isolated bacteria (usually from the patient's sputum) are cultured in presence of a drug. If sufficient growth is observed in a preset timespan, the bacteria are diagnosed as resistant. Unfortunately, this process can take weeks, a period during which the patient is treated with ineffective drugs allowing the resistant case to further spread, and potentially develop resistance to additional drugs.

Drug resistance in *Mycobacterium tuberculosis* (*M. tuberculosis*), the causative agent of TB, commonly emerges as a result of a point mutation in specific genes. This knowledge has been exploited in the development of molecular diagnostics as a rapid alternative to growth-based methods. Molecular testing, however, has an important disadvantage in that it can only detect resistant mutations that the platform was designed to detect. Bacteria that harbor uncommon mechanisms of resistance, will therefore, escape detection. As a result, the sensitivity in detecting resistance can suffer with increased prevalence of uncommon mechanisms of resistance. It is well documented that mono isoniazid resistant (INH$^R$) bacteria often harbor such mechanisms, escape detection, and develop into MDR-TB.(1, 2) As a result, incorporation of all resistance conferring mutations is critical for comprehensive molecular detection. For this reason, we set out to catalog all resistance conferring mutations and estimate global prevalence of common (canonical) and uncommon mutations that confer INH resistance. In doing so, we used the search criteria used by the previous systematic review by Seifert et al. (3), so the two study's results can be compared and temporal changes in prevalence can be estimated. While Seifert et al. surveyed articles published between year 2000 and August 2013, this study continued that survey from September 2013 until December 2019.

Previous studies have shown that mutations in *katG*, *inhA*, *fanG1*, and the *oxyR'-ahpC* intergenic region confer INH resistance.(3, 4) *katG* codes for a catalase-peroxidase that activates INH. Mutations in *katG*, in particular at codon 315, are commonly observed to cause resistance to INH. The second most frequently observed mechanism of resistance is through mutations in the promoter of the gene *inhA*, in particular at position -15. (3)(4) This leads to overexpression of the gene which results in removal of the drug from the bacterial cell. (3) Mutations in the coding region of the *inhA* gene also have been associated with resistance. (3)(4) Mutations in the *oxyR'-ahpC* intergenic region may help alleviate the fitness cost of the loss of KatG activity in many resistant isolates by increasing the expression of *ahpC*. (4) Finally, a synonymous mutation in *fabG1*, L203L (CTG203CTA), has been shown to cause INH resistance through the creation of an alternative promoter for *inhA*, and hence causing its overexpression. (5, 6)

In this review, we catalog the mutations reported as conferring INH resistance in the three genes and the intergenic region between September 2013 and December 2019, report their individual frequencies, estimate the sensitivity of a molecular platform in detecting INH resistance, solely based on common mutations, as well as based on all mutations reported in this review. We finally, compare the results of this review with that of the previous systematic review to see how molecular epidemiology of INH resistance has changed in the last five years.

## METHODS

### Literature Search

A search in PubMed was conducted on all peer-reviewed publications evaluating mutations in *katG*, *inhA*, and *oxyR-ahpC* intergenic region in INH$^R$ clinical isolates of *M.tuberculosis*. In order to continue the previous systematic review (3) presented by Seifert et. al., I used the same search term used by that study: (isoniazid OR inh) AND (resistance OR resistant) AND (mutations OR mutation) AND tuberculosis. The search was limited to studies published between September 2013 and December 2019 in order to avoid the aggregation of strains that were reported by Seifert et. al. That study reported articles published through August 2013. (3)

### Study Selection Criteria

Studies were included if: 1) written in English; 2) presented original data; 3) used clinical strains of *Mtb*; 4) described the phenotypic DST method used as reference standard; 5) at a minimum reported frequency of canonical mutations (in *katG* codon 315 and *inhA* promoter positions -8 and -15); 6) included individual level amino acid mutation data. Mutations in the putative regulatory regions or promoter region were included if available. Studies that performed DST on solid or liquid media were included, as long as cut-off concentrations were clearly defined. A range of genotypic testing platforms were allowed: Whole Genome Sequencing (WGS) (PacBio RS), Line Probe Assay (LPA) using GenoType® MTBDRplus (Hain Lifescience, Nehren,

Germany), PCR, GeneXpert (Cepheid, Sunnyvale, CA), and proprietary platforms (only if confirmatory data from a secondary platform was available).

The primary goal of this survey was to reassess the prevalence of canonical mutations. As such, any report that did not provide detailed co-mutation (e.g. isolates harboring both *katG*315 and *inhA* -15) counts was excluded since this resulted in overestimation of the prevalence of these mutations.

*Data accuracy*: To assure the accuracy of the data the following step-wise methodology was taken:

1. This process of mutation curation was repeated twice independently: Each manuscript was evaluated independently twice, once in February 2020 and a second time in March 2020.
2. Discordance resolution: The manuscripts with discordant counts (between the two curation rounds) were assessed for a third time to resolve the discordance.
3. Parity error check: Two methods were used to detect frequency errors:
    a. Per-locus parity: If the sum of the mutations and wild-type observations reported for any locus did not add up to the total isolate count reported in the text, the article was excluded.
    b. Isolate count parity: If the total number of mutant and WT isolates reported in the tables did not add up to the total isolate count reported, the article was excluded. Not reporting of co-mutations, is the common cause of this.

**Data Acquisition**

The following data was extracted from articles that met the inclusion criteria: author names, publication year, PubMed ID, DOI, title of the paper, total number of resistant isolates (reported and calculated), total number of isolates harboring each mutation/combination, geographic location, method of determining DST, and the method for detecting mutations. Each paper was examined for individual mutations and combination of mutations in *katG*, *inhA*, *fabG*1, and *oxyR-ahpC* intergenic region. Mutations in the same locus but with a different change was reported separately. The frequency of multiple mutations harbored by the same isolate (e.g. *katG* 315 and *inhA* -15) was reported as a combination, distinct from the frequency of each mutation appearing as a stand-alone mutation in isolates.

In this manuscript, we report the locus of each mutation with respect to *Mtb* H37Rv genome (Accession number NC_000962.3).

**Quality Assessment**

For each mutation reported, the reported reference amino acid was compared to the published H37Rv sequence (Accession number NC_000962.3). Mutations reported with a reference amino acid discordant with that of H37Rv were excluded from our analysis.

**Statistical Analysis**

Per-mutation analysis: The total number of isolates harboring each distinct mutation (or combination thereof) was calculated across all included articles and divided by the total number of resistant isolates included in all articles to estimate the diagnostic sensitivity of the mutation.

Per-locus analysis: The total number of resistant isolates that harbored a mutation in each distinct locus was calculated across included articles (regardless of regardless of the type change). The total at each locus was then divided by the total number of resistant isolates included in all articles to estimate the diagnostic sensitivity of the locus. Since the purpose of this analysis is to indicate the importance of inclusion of the locus in diagnostics, combination of mutations that included the locus were also included in this analysis.

Per-region/gene analysis: To estimate the diagnostic significance of a gene (*katG*, *inhA* including its promoter) or a region (*oxyR'-ahpC* intergenic region), the total number of resistant isolates reported to harbor a mutation in each region was calculated across the included articles and divided by total number of resistant isolates.

## RESULTS

### Description of Included Studies

Our search through PubMed Medline using the terms indicated resulted in 509 articles published between September 1, 2013 and December 31, 2019. Out of the 509 potential studies, 52 studies met all inclusion criteria. (1, 7–59) A PRISMA flow diagram in Figure 1 illustrates the breakdown of number of articles excluded because of the specified criterion.

In total, 5792 $INH^R$ isolates were reported by the 52 included articles. Table 1, presents the breakdown per region. Overall, nine different methods were used in the 52 included studies to detect mutations. The most prevalent method was PCR followed closely by WGS, and

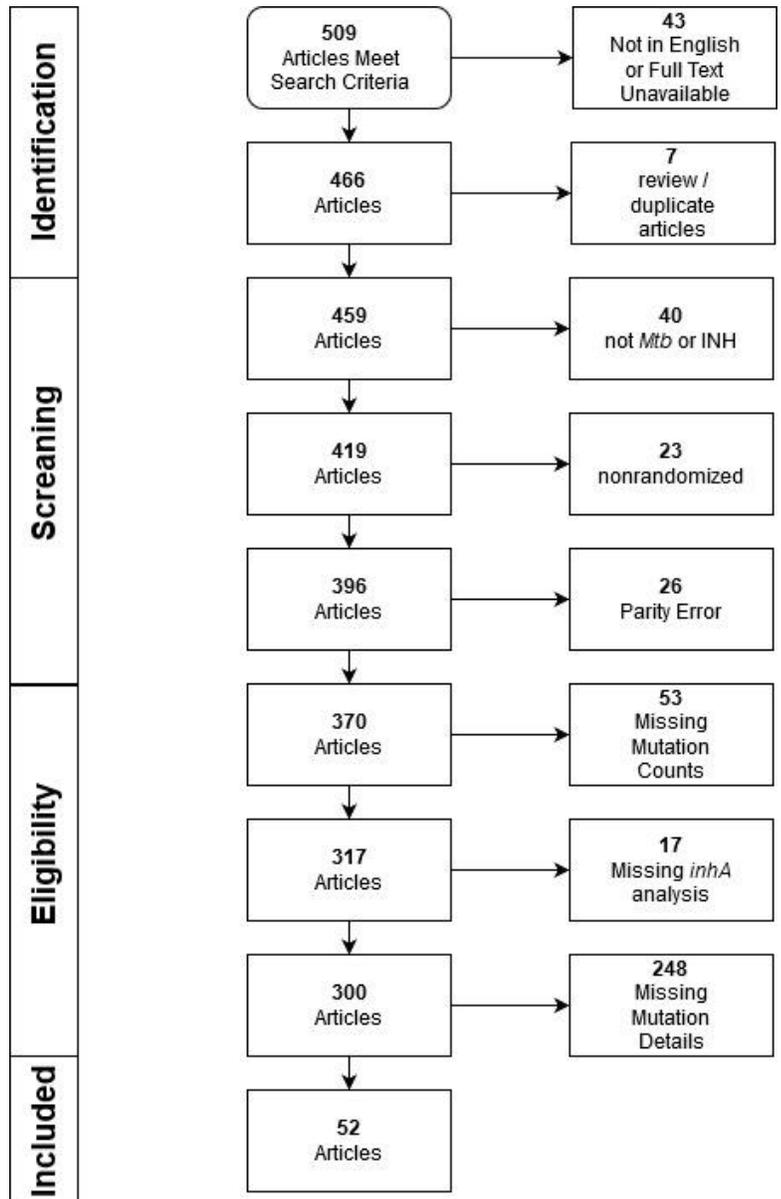

**Figure 1.** PRISMA Flow Diagram. Counts for exclusion criteria. Pubmed search term: (isoniazid OR inh) AND (resistance OR resistant) AND (mutations OR mutation) AND tuberculosis

**Table 1.** Breakdown of INH$^R$ isolate counts per region of the world. Percentages reflect the percentage of each count with respect to the total (5792) INH$^R$ isolates included in this study.

|  | Africa (1074=18.5%) | | | Asia (4110=71.02%) | | | | Europe (235=4.05%) | | | Americas (373=6.43%) | |
| --- | --- | --- | --- | --- | --- | --- | --- | --- | --- | --- | --- | --- |
| Region | South | North | East | West | Central | South | East | West | South | East | South | North |
| Count | 415 | 39 | 620 | 141 | 348 | 1980 | 1641 | 51 | 8 | 176 | 344 | 29 |
| Percentage | 7.15 | 0.67 | 10.68 | 2.43 | 6.00 | 34.11 | 28.48 | 0.88 | 0.14 | 3.03 | 5.93 | 0.50 |

MTBDRplus. The three methods combined for sequencing over 82% of included isolates. We separated LPA and PCR platforms with the distinction that PCR is the category of LPA platforms designed inhouse (typically in academic settings). Among such studies, we only included those that had reported the PCR primers for each of the canonical mutations. Table 2, presents the full list of nine methods and the number of isolates they each sequenced. As compared to the previous review(3), perhaps the most notable change in sequencing methodology has been the increase in use of WGS. The infrequent use of GeneXpert was also surprising.

**Table 2.** Number of isolates stratified by the genotyping platform.

| Method | Count |
| --- | --- |
| PCR | 1643 |
| WGS | 1580 |
| GenoType MTBDRplus | 1566 |
| LPA | 422 |
| GenoTypeCM Mycobacterium Assay | 157 |
| QMAP | 144 |
| TB-BIOCHIP | 135 |
| HRM /PCR | 131 |
| GeneXpert | 14 |

Mutations in *katG*, *inhA* (promoter and gene), and *oxyR-ahpC* intergenic region were analyzed and are reported in the following section.

*Mutations in KatG.* Two sets of mutations were excluded from our analysis: synonymous mutations that do not cause any change in the protein, and mutations in codon 463 since they appear abundantly among INH$^R$ as well as INH$^S$ isolates. As depicted by Figure 2, 73 (or 9.8%) of the 740 amino acids in KatG harbored a mutation in at least one isolate. Additionally, one study reported a resistant isolate with deletion of the entire *katG* (32), while a second reported the deletion of *katG* as well as a substitution of C-52T in the *oxyR-ahpC* intergenic region of a resistant strain (52). Deletion of *katG* has been shown to cause resistance to isoniazid. In total, ignoring codon 463, 4,220 (74.93%) resistant isolates harbored a nonsynonymous mutation in *katG*. While mutations were observed across the gene, the range between codons 235-350 could be considered a relative hot spot, with 4,356 (75.05%) resistant isolates harboring a mutation in this region. Codon 315 in *katG* of 4,100 resistant isolates harbored a mutation, providing for a diagnostic sensitivity of 72.80%. As expected, this was the locus with the highest frequency of mutations. Among the varieties of mutations observed in this codon, *katG* S315T (AGC-ACC) was the most common variety and was harbored by 3,673 (or 65.22%) of resistant isolates.

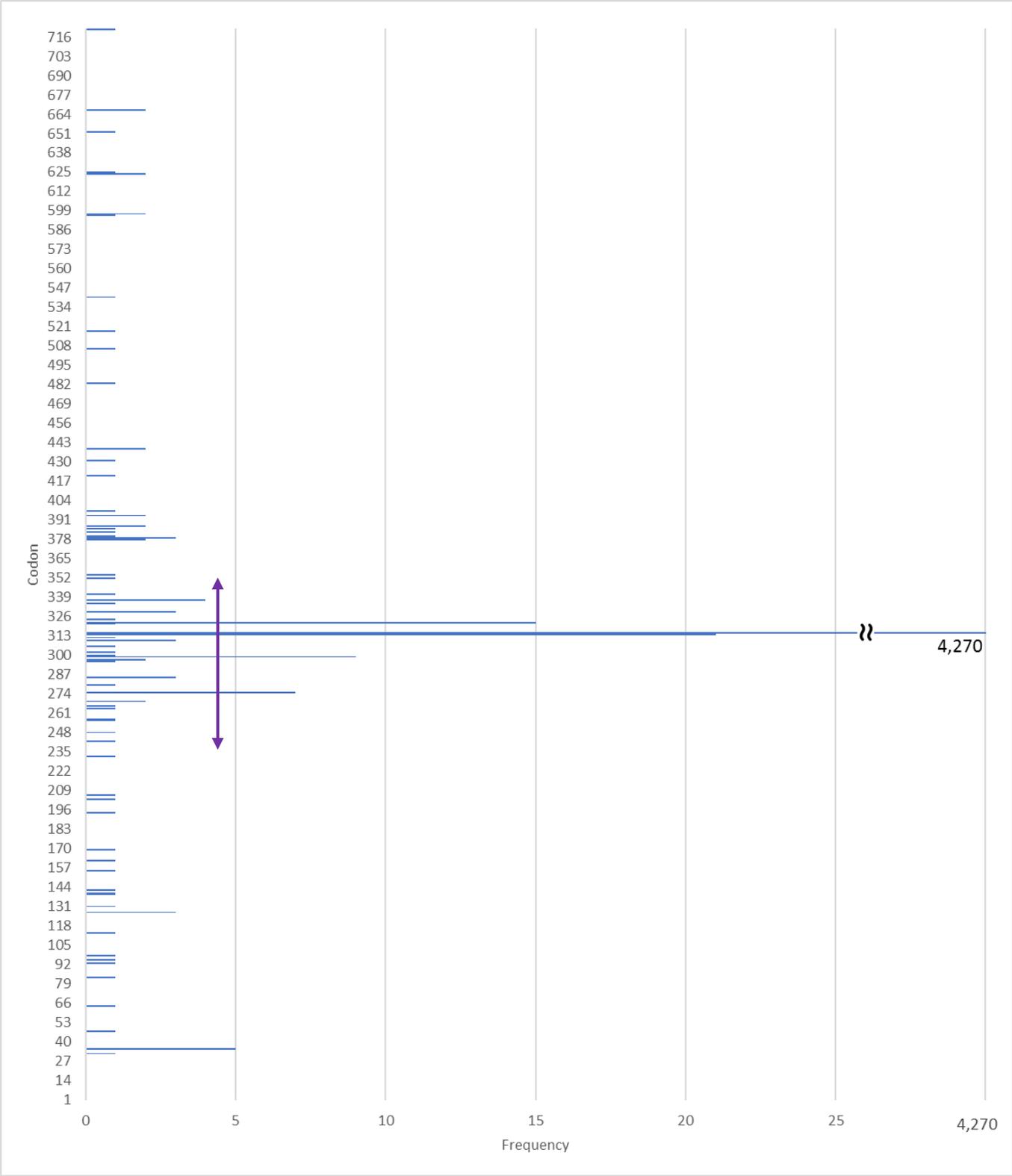

Figure 2. Frequency of mutations in *katG* observed in INH[R] clinical isolates.

*Mutations in inhA promoter and open reading frame.* The promoter region of *inhA*, was mutated in 966 (17.15%) of the resistant isolates. Of these, 471 (8.36%) resistant isolates also had a mutation in KatG, while five had two mutations in *inhA* or its promoter. Mutations in 16 loci were reported in *inhA* or its promoter. Table 1 lists the loci, the number and percentage of resistant isolates that harbored a mutation in each position. The most prevailing *inhA* promoter locus was -15, with 776 (13.78%) resistant isolates harboring a mutation at this locus. At this locus, the most prevalent mutation was *inhA* C-15T. Among resistant isolates, 454 (8.06%) harbored this mutation.

**Table 1:** Locus describes the position of the mutation in the promoter (negative loci) or within (positive loci) *inhA* gene. Positive loci are codon numbers while negative loci are nucleotide positions within the promoter of the gene.

| Locus | -94 | -34 | -17 | -16 | -15 | -10 | -9 | -8 | -6 | -3 | 9 | 11 | 71 | 94 | 113 | 231 |
|---|---|---|---|---|---|---|---|---|---|---|---|---|---|---|---|---|
| Count | 1 | 16 | 33 | 14 | 787 | 3 | 1 | 106 | 1 | 6 | 2 | 1 | 1 | 3 | 1 | 3 |
| % | <0.1 | 0.3 | 0.6 | 0.2 | 13.6 | 0.1 | <0.1 | 1.8 | <0.1 | 0.1 | <0.1 | <0.1 | <0.1 | 0.1 | <0.1 | 0.1 |

*Mutations in oxyR-ahpC intergenic region.* Mutations in the 106bp intergenic region between the pseudogenes *oxyR* and the gene *ahpC* (H37Rv genome positions 2,726,088 to 2,726,192) are known to be associated with resistance, although Vilcheze et. al. (4) suggested that mutations in this region may not cause resistance but rather compensate (through overexpression of *ahpC*) for loss of peroxidase function caused by *katG* mutations. Regardless of the role, the association of mutations in this region make them viable candidates for molecular diagnostics. Importantly, several studies did not report mutations in this region due to the limitations of the molecular platform used for detection of mutations (e.g. GeneXpert). As such, the frequencies reported for mutations in this region, underrepresent the frequency of such mutations.

In total, 113 (2.01%) resistant isolates harbored a mutation in this region or in *ahpC* coding region (2 isolates). Of the 113, 31 (27.43%) isolates also harbored a KatG mutation, while 14 (12.39%) harbored an *inhA* (promoter or InhA) mutation. Seven (6.19%) of the 113, harbored a mutation both in KatG and *inhA* (promoter or InhA). Importantly, 75 (66.37%) of the 113 resistant isolates, did not harbor a mutation in KatG, or *inhA* (promoter or InhA).

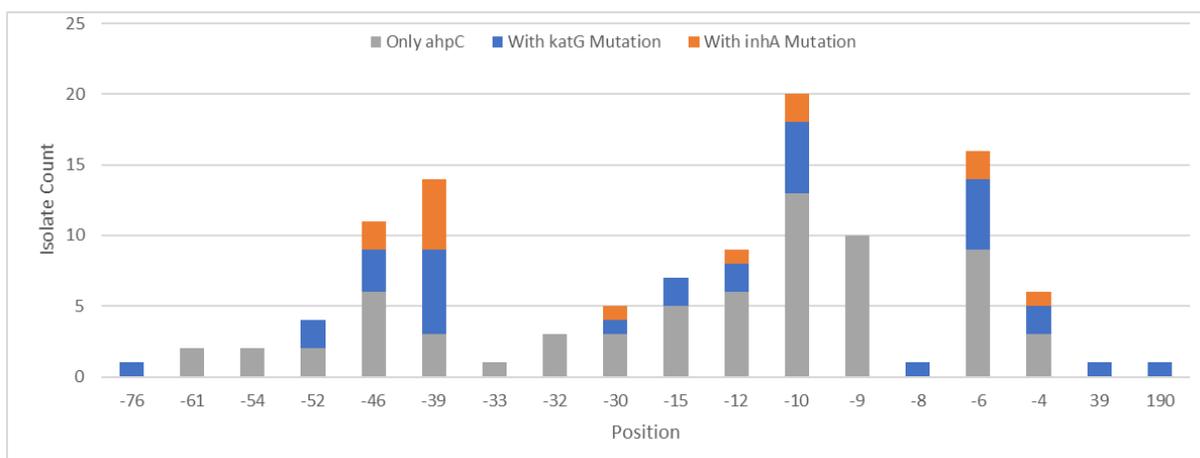

**Figure 3:** Mutations in intergenic region between the genes *oxyR* and *ahpC* (106bp in H37Rv genome positions 2,726,088 to 2,726,192). Negative positions are nucleotide positions relative to the start of the gene *ahpC* on the positive strand while positive positions are codon numbers in the gene's ORF. Blue bards indicate isolates that also harbor KatG mutations, while orange bars indicate isolates that also harbor *inhA* (promoter or gene) mutations. Grey bars indicate isolates that harbor no mutations in *inhA* (promoter or gene) or in KatG. Seven isolates had a mutation in *katG* and *inhA*.

# DISCUSSION

Molecular diagnostics is rapid, often cheaper, and at times more portable, promising to bring rapid testing to bedside and even into the community at a low cost. It does however take testing one more step away from clinical outcomes. While phenotyping was designed to be used to predict treatment outcome for a given regimen or drug, molecular platforms are often designed and tested as a proxy for the lengthier phenotypic testing. As a result, detailed knowledge about the performance of molecular testing and its assumptions is essential.

One of the dangers of molecular diagnostics is that selection that it could impose on to the molecular epidemiology of drug resistant TB. A broadscale application of molecular platforms that aim to detect only the canonical mutations, increases the risk of lowering the incidence of isolates that harbor canonical mutations, since they are readily detected and appropriately treated, while those with uncommon mechanisms of resistance will escape detection and further spread. I will refer to this effect as the "unnatural" selection for uncommon mechanisms. This is of particular importance for isoniazid resistance since its canonical mutations are well defined (e.g. *katG*315, *inhA*-15, *inhA*-8) while quite a few other mutations are also known to cause resistance and frequently appear without the presence of canonical mutations. Such strains are likely to escape detection by molecular testing and further spread.

This study was designed with two aims in mind: 1) to catalogue mutations associated with INH resistance; 2) to assess the prevalence of canonical mutations; 2) estimate their global prevalence and co-occurrence and their utility in molecular diagnostics. For aim 1) I include a summary of all the mutations reported in 52 included manuscripts. For aim 2, I report all mutations' frequencies and compare these results to that of the previous survey by Seifert et al.(3) who reviewed articles published between January 2000 and August 2013. While this comparison is suboptimal, any notable differences provide additional evidence for our hypothesis that frequent global and regional reevaluation of sensitivity and specificity of molecular platforms are essential to prevention of emergence of MDR-TB.

Canonical mutations: As defined for this manuscript, we have considered mutations that appear in codon 315 of *katG* and promoter mutations at -15 and -8 positions for *inhA* to be canonical. All molecular platforms reported in this survey have targeted these positions for mutations. The comparison of the prevalence that we observe as compared to those observed by Seifert et al.(3) is presented in Table 4.

**Table 4:** Frequency of canonical mutations (September 2013 – December 2019) in this study as compared to the previous systematic review by Seifert et al.(3) ending in August 2013. *Combining frequencies reported for *inhA*-8 and *mabA*-8 by Seifert et al.(3).

| Mutation | *katG* 315 | | *inhA* -15 | | *inhA* -8 | |
|---|---|---|---|---|---|---|
| **Study** | This Study | Seifert et al. | This Study | Seifert et al. | This Study | Seifert et al. |
| **Frequency** | 72.8% | 66.2% | 14.0% | 19% | 1.9% | 2.25%[*] |

As it can be seen, the frequency of both canonical mutations in the promoter region of *inhA* are notably lower than the previous systematic review while the frequency of *katG*315 is higher in our review. The cumulative frequencies of these mutations are compared in Table 5. Importantly, Seifert et. al. reported mutations in *mamA* and *fabG1* separately. Since these two locus tags refer to the same gene, frequencies reported for this gene appears lower than it should be. Additionally, since *fabG1* and *inhA* belong to the same operon, promoter loci are reported with respect to the beginning of the operon which is the beginning of *fabG1* coding region. For example, *fabG1*-8 and *inhA*-8 tags point to the same locus, namely eight positions upstream of *fabG1*. However, since the authors have reported these separately, the frequencies for these loci appear to be lower than they should have been. As such, if we combine the frequency reported for *inhA*-8 (1% in Table 4) and *mabA*-8 (1.25% Supplemental Table 4), the total frequency of this locus would be 2.25%.

**Table 5:** Cumulative frequency of canonical mutations (September 2013 – December 2019) in this study as compared to the previous systematic review by Seifert et al.(3) ending in August 2013.

| Mutation | *katG* 315 + *inhA* -15 | | *katG* 315 + *inhA* -15 + *inhA* -8 | |
|---|---|---|---|---|
| **Study** | This Study | Seifert et al. | This Study | Seifert et al. |
| **Frequency** | 81.4% | 79.9% | 81.9% | Not reported |

*Mutated loci in katG*: Seifert et. al. reported observing at least one mutation in 202 *katG* codons. In this study, we report 73 such loci. Of these, 32 loci are new observations as compared to the previous systematic review. Forty-one codons were common between the two studies. It is a conclusion of this study that the subset of 41 codons shared between the two studies is a more reliable basis for molecular diagnostics rather than those that appear at a slightly higher frequency in one study. Several reasons could explain this, including the possibility of clonal expansion in an outbreak that might artificially inflate the numbers locally at a given time, but carry less significance globally or over a longer span of time. A good example is the set of four *katG* codon recommended by Seifort et al. for diagnostics: 309, 311, 315, and 316. These codons were the most frequently mutated codons in their study, where a mutation was observed in 36, 27, 4059, and 27 INH$^R$ isolates respectively. Of this set of four, only one, codon 315, was observed to harbor a mutation in this study. The remaining three would not increase the sensitivity of molecular diagnostics in our set of 5792 INH$^R$ isolates.

*Mutated loci in inhA*: In total, Seifert et al.(3) report a mutation in 44 loci in the promoter or coding region of *inhA*, while we report only 16 such loci in our isolates. Of these, mutations in 10 loci are new observations as compared to the previous systematic review. Only six (five promoter and one coding region) loci were common to both studies: -34, -17, -15, -9, -8, and 143. As in the case of *katG*, we recommend these loci for diagnostics. This is mostly inline with Seifert et. al.'s recommendation with the exception that we have replaced the -47 locus with codon 143.

An argument could also be made for inclusion of both loci.

*Mutated loci in oxyR'-ahpC*: Seifert et. al. reported observing a mutation in 32 loci in this intergenic region or in *ahpC* coding region while we report observing a mutation in 18 loci. Of these mutations in seven loci are new observations as compared to the previous systematic review. Eleven loci appeared in both studies: -54, -52, -46, -39, -32, -30, -15, -12, -10, -9, and -6. Importantly, while *aphC* coding loci appeared in the two studies, none were common to both.

In summary, in addition to the 14.1% of isolates that harbor a yet to be detected mechanism of resistance, those that carry a known mechanism but one that is not included in current molecular diagnostics, are a source of deep concern. With broad implementation of molecular diagnostics, the mechanisms of resistance that are included in molecular diagnostics will be identified and eradicated while the remaining resistant strain will escape detection and continue infect others. This will result in an "unnatural" (man-made) selection that will result in lower regional detection rates and increased prevalence of uncommon mechanisms of resistance. To avoid this, systematic reviews such as this need to be annually repeated and the information discovered be be routinely included in molecular diagnostics.

## ACKNOWLEDGEMENTS

I would like to acknowledge Dr. Michael Botte's (Scripps Clinic) mentorship in all aspects of medical practice including surgery. His mentorship played an important role in maintaining my interest in medicine and hence the perseverance that was needed for completion of this work. I also would like to acknowledge Dr. Wael Elmaraachli (UCSD Division of Pulmonary Medicine) for his vast subject area expertise and his mentorship in pulmonary medicine, in particular in the treatment of tuberculosis, complications relating to drug resistance.